# Chemical exfoliation of MoS$_2$ leads to semiconducting 1T′ phase and not the metallic 1T phase

Banabir Pal,[1] Anjali Singh,[2] Sharada. G,[1] Pratibha Mahale,[1] Abhinav Kumar,[1] S. Thirupathaiah,[1] H. Sezen,[3] M. Amati,[3] Luca Gregoratti,[3] Umesh V. Waghmare,[2] and D. D. Sarma[1*]

[1] Solid State and Structural Chemistry Unit, Indian Institute of Science, Bengaluru 560012, India

[2] Jawaharlal Nehru Centre for Advanced Scientific Research, Jakkur, Bengaluru 560064, India

[3] Elettra - Sincrotrone Trieste S.C.p.A., SS14 - km 163,5 in AREA Science Park, 34149 Basovizza, Trieste, Italy

Abstract: A trigonal phase existing only as small patches on chemically exfoliated few layer, thermodynamically stable 1H phase of MoS$_2$ is believed to influence critically properties of MoS$_2$ based devices. This phase has been most often attributed to the metallic 1T phase. We investigate the electronic structure of chemically exfoliated MoS$_2$ few layered systems using spatially resolved (≤120 nm resolution) photoemission spectroscopy and Raman spectroscopy in conjunction with state-of-the-art electronic structure calculations. On the basis of these results, we establish that the ground state of this phase is a small gap (~90 meV) semiconductor in contrast to most claims in the literature; we also identify the specific trigonal (1T′) structure it has among many suggested ones.

2D transition metal dichalcogenides have emerged as a viable alternative to graphene with extraordinary properties and potential applications[1-3]. Molybdenum disulfide ($MoS_2$) is undoubtedly the preeminent member in the family for applications in transparent and flexible electronics.[4-13] While the usual crystallographic form of $MoS_2$ is the hexagonal 1H phase, $MoS_2$ exhibits a number of trigonal polymorphic forms[14-16], such as 1T, 1T′, 1T″ and 1T‴ (see Fig. 1), distinguished by small distortions. These metastable states can be kinetically formed as small patches embedded in the majority 1H phase during chemical exfoliation, which is an attractive, easily scalable route to obtain one or few layer $MoS_2$ in substantial quantities[17]. Even mechanically exfoliated $MoS_2$ may have small quantities of these metastable forms, influencing its material and device properties.[18] The stable 1H form has been extensively studied and its electronic properties are well understood as a semiconductor with a large (1.9 eV) band gap. Unfortunately, electronic structures of different polymorphic $MoS_2$ are not known, though it may potentially limit or enhance the applicability of two dimensional $MoS_2$ devices by its presence within 1H $MoS_2$ samples. It has been generally assumed that the metastable phase is of 1T form and metallic in nature[19-23]. This presumed metallic nature is considered to be the cause of some novel beneficial device properties as well. For example, metallic 1T phase is believed to be responsible for the very high energy and power densities in supercapacitors[21] and also for the remarkably high hydrogen evolution reaction efficiency[22] achieved using chemically exfoliated $MoS_2$. Whatever little is known of electronic structures of metastable phases is primarily from theoretical calculations that present contradictory views, ranging from being metallic (1T phase) to normal insulator (for 1T′ phase[24]) or even ferroelectric insulator (1T‴ phase[25]). Surprisingly, direct structural investigations, based mostly on TEM and STM, also lack any agreement between different reports with the crystal structure of the metastable patches of $MoS_2$ being variously ascribed to the 1T form,[19,26] the distorted 1T″ form with a 2a×2a superstructure,[27,28] the 1T′ form[29,30] with a zigzag chain-like clustering of Mo atoms, and also the distorted 1T‴ with a √3a×√3a superstructure[29,31] where a trimerization of Mo atoms takes place (see Fig.1). Thus, the hope to understand the true electronic structure of this important phase of $MoS_2$ via an experimental determination of its geometric structure in conjunction with electronic structure calculations has not been realized so far.

The main difficulty in experimentally probing this metastable phase of $MoS_2$ is that it exists only in small patches in the 1H matrix of few layer $MoS_2$. Photoemission spectroscopy is the only direct probe of the electronic structure due to its inherent extreme surface sensitivity. However, the usual practice of photoemission spectroscopy does not have the required spatial resolution to enhance the relative contribution from the microscopic patches of the metastable phase. Therefore, we have carried out spatially resolved photoemission investigation with a ~120 nm photon beam diameter to directly determine the electronic structure of this metastable phase of $MoS_2$ and conclusively establish that this elusive phase is actually a small gapped (~90 meV) semiconductor in sharp contrast to the dominant belief of it being metallic. We use state-of-the-art electronic structure calculations to provide evidence that this phase corresponds to the 1T′ structure, supported by micro-Raman experiments.

The minority phase of $MoS_2$ was extensively stabilized on conducting indium doped tin oxide (ITO) substrates using the well-known organolithium route.[17] The mechanically exfoliated pure sample of $MoS_2$ and the one after chemical treatment to stabilize the metastable polymorphic form are referred to as A and B, respectively in this manuscript.

Raman scattering is a powerful and sensitive tool to detect presence of different phases of any given material and has been extensively used for low dimensional materials in recent times, providing vibrational fingerprints for different phases. Therefore, we characterized samples A and B using a micro-Raman probe with a spatial resolution of ~1 μm, with only three representative spectra shown for each sample. Raman spectra of the pure 1H sample (Fig. 2a) exhibit two peaks at 383 cm$^{-1}$ and 408.5 cm$^{-1}$ due to $E^1_{2g}$ and $A_{1g}$ modes, respectively, consistent with earlier reports.[32] For sample B (Fig. 2b), we observe three additional peaks at 156 cm$^{-1}$, 227 cm$^{-1}$ and 330 cm$^{-1}$, referred to as $J_1$, $J_2$ and $J_3$ modes respectively, in the literature[21,32] and associated with the formation of the metastable trigonal phase. We could not find any spot on this sample with only signals of the metastable phase with no signature of the 1H phase, clearly indicating that the patch size of this chemically induced MoS$_2$ phase is smaller than 1 μm.

Scanning Photo Electron Microscopy (SPEM) measurements were performed to understand the electronic structure of the chemically exfoliated system. Fig. 3a shows a typical photoelectron spectrum in the Mo 3$d$ region with two narrow peaks at binding energies of 229.3 eV (Mo 3$d_{5/2}$) and 232.4 eV (Mo 3$d_{3/2}$) and a broad, low intensity feature at ~226.5 eV due to S 2$s$ contributions. We compare this with a typical spectrum obtained from the sample B. Clearly, peaks in the spectrum of the mixed phase are specifically broadened on the lower binding energy side. Since the intensity of Mo 3$d_{5/2}$ peak from pure 1H appears almost entirely between 228.7 and 230.0 eV in Fig. 3a, the additional intensity from sample B between 228 eV and 228.7 eV must arise from polymorphic forms of MoS$_2$ other than 1H. This shift in the binding energy of the metastable form allows us to map its presence in the form of an image by scanning the sample through the focused photon beam and plotting the relevant Mo 3$d_{5/2}$ intensity. In the SPEM imaging mode, first we carried out a detailed mapping of Mo $3d_{5/2}$ intensity over the entire 228-230 eV binding energy window, covering contributions from both phases and thus, imaging the distribution of MoS$_2$ without any reference to its polymorphic forms on the ITO substrate as shown in Fig. 3b. Then, we plot in Fig. 3c an image of the ratio of intensities corresponding to energy windows, I and II, shown in Fig. 3a, corresponding to Mo $3d_{5/2}$ intensities arising primarily from the metastable phase and the 1H phase, respectively. The contrast in the intensity ratio, I/II, being independent of topographic features[33], reveals the relative abundance of the metastable phase in the sample B with the dark blue regions corresponding to the metastable MoS$_2$ rich region. We point out that almost all spots imaged in Fig. 3c contain signals of both the 1H and the metastable phase, explicitly checked by recording Mo $3d_{5/2}$ spectra from over 30 different spots; there were only few spots that corresponded to the pure 1H phase and no spot that had contribution only from the metastable phase. Coupled with the fact that there is a considerable intensity contrast even with the present photon spot size, this implies that the typical size of metastable patches is in the order of ≤120 nm, but not very much smaller.

We identified a region with the largest contribution from the metastable phase, marked by the rectangular window in Fig. 3c; the intensity of the Mo $3d_{5/2}$ from the metastable phase was essentially uniform over this region. Thereafter, we carried out a detailed spectroscopic investigation with the photon beam focused at the center of this window to maximize contributions from the metastable phase that we are interested in. The Mo $3d$ spectrum obtained from this spot is shown in Fig. 3d. We have decomposed this spectrum in terms of contributions from the 1H phase and the metastable phase using a least squared-error approach with the spectral feature of the 1H component as determined from measurements of sample A and that for the metastable phase approximated by a Lorentzian line convoluted with a Gaussian function with full width at the half maximum, FWHM, representing the life-time and the resolution broadenings. The resulting components, also shown in Fig. 3d, establish the dominance of the metastable phase at this spot on the sample, with the area ratio (~2.8) of the two components providing a quantitative measure of the relative abundance of the two phases. We also note that the electronic binding energy difference (~0.7 eV) between the two components is in agreement with earlier publications.[18, 23]

There have been suggestions in the literature that the presence of Li ions on such chemically exfoliated samples and consequent charge doping may influence the formation of a specific metastable state. Therefore, we scanned carefully the binding energy region corresponding to Li 1$s$ level with a high counting statistics and found no evidence of any presence of Li in our samples. This shows that the sample preparation method, that involves repeated washing of the chemically exfoliated samples first with hexane and then with water is effective in removing all traces of Li. We also note that the expected charge doping in presence of Li would tend to drive the system towards a metallic state, which is also in contradiction with our valence band results, discussed next.

We show the valence band spectrum of the pure 1H phase obtained from sample A in Fig. 4a with features appearing at 1.8, 3.3, 4.5, 5.2 and 6.5 eV binding energies arising from hybridized Mo 4$d$ and S 3$p$ states. Electronic structure calculated within a local density approximation of DFT without or with the inclusion of the spin-orbit coupling did not provide an accurate description of our experimental results, significantly underestimating the energy separation between dominantly

Mo 4*d* (1.3 eV) and S 3*p* (3.3 eV) peaks. The inclusion of on-site Coulomb interaction of Mo 4*d* states within the standard LDA + U approach made this comparison worse. Noting the importance of including correlation effects on all sites for a proper agreement with photoemission valence band spectra[34, 35], we used the GW approximation[36] to calculate the electronic structure of 1H MoS$_2$, in a periodic supercell to simulate a 2D MoS$_2$ sheet with a vacuum of 10 Å between the periodic images, within Vienna ab initio simulation package (VASP) [37] with GW version of PAW[38,39] potentials, which are expected to provide improved scattering properties at high energies[40].

To compare the calculated partial DOS (Fig. 4a) to the experimental spectrum, we took into account the relative cross-sections of Mo 4*d* and S 3*p* states at the photon energy used[41]. Fig. 4a shows the sum of these cross-section weighted partial DOS after the usual Lorentzian and Gaussian broadenings. The agreement between the calculated result and the experimentally obtained spectrum is remarkable, firmly establishing both the need and the efficacy of the GW calculation scheme in capturing the electronic structure of MoS$_2$.

Fig. 4b shows the valence band photoemission spectrum (inset) and only the near- Fermi energy ($E_F$) region (main frame) from the same spot of sample B, in comparison to the near-$E_F$ spectrum of the pure 1H phase, with all spectra exhibiting negligible intensity at $E_F$, implying a nonmetallic state. Fig. 4b also shows that the spectral intensity of the MoS$_2$ 1H phase is negligible down to ~0.5 eV binding energy, indicating a sizable gap. The existence of this large gap in the photoemission spectrum of the pure 1H sample allows us to understand electronic states close to $E_F$ in the metastable phase, as the spectrum from sample B within the range of binding energies smaller than ~0.5 eV is contributed only by the metastable phase. In the photoemission spectrum close to $E_F$ simulated with model densities of states (DOS) broadened by resolution and lifetime effects, the gap between the top of the valence band and $E_F$ is ~90 meV, which clearly defines the metastable $MoS_2$ phase generated by the chemical exfoliation of 1H $MoS_2$ as an insulator in sharp contrast to the dominant assumption of it being the metallic 1T phase.

To go beyond the direct experimental evidence of a small gap in the metastable state and understand its electronic and geometric structure, we first extract the valence band spectrum of the metastable phase alone by subtracting the contribution of the 1H phase in the total spectrum recorded experimentally. This can be reliably performed, since we already know the relative abundances of the metastable and the 1H phases from exactly same spot on the sample from the two distinct Mo $3d$ core level contributions arising from the two phases, as shown in Fig. 2d; this relative abundance of the two phases and the knowledge of the valence band spectrum of the pure 1H phase (Fig. 4a) allow us to subtract accurately the 1H spectral contribution (triangles in Fig. 4c) from the total spectrum (diamond symbols in Fig. 4c). In addition, we also remove the low intensity, featureless contribution (line in Fig. 4c) to the total valence band spectrum arising from the ITO substrate due to the extreme thinness of the $MoS_2$ sample; this is achieved quantitatively by recording the In $4p$ core level and the valence band spectra of the pure substrate without any $MoS_2$ sample as well as the In $4p$ core level spectrum from the substrate with the sample B on top of it. The resulting valence band spectrum of only the metastable state, shown as circles in Fig. 4c and also in Fig. 4d, appears to be distinctly different from that of 1H phase. It is interesting to note that this spectrum lacks characteristic sharp features of the 1H phase, pointing to a lower symmetry of the metastable phase or in other words, some distortions in the lattice.

The electronic structure reveals the underlying lattice responsible when analyzed in conjunction with a reliable theoretical approach to describe the electronic structure. Adopting the GW approach, established to be accurate for the 1H phase, for all possible trigonal metastable phases of MoS$_2$, namely 1T, 1T′ and 1T‴, we find that (1) the 1T structure is metallic, as already known[19] and therefore, is inconsistent with experimental observations (Fig. 4d); (2) both 1T′ (2a×a) and 1T‴ (√3a×√3a) superstructures lead to insulating ground states with band gaps of ~45 and ~500 meV, respectively. However, it turns out that the DOS and the partial DOS of only the 1T′ superstructure, combined with the same matrix elements and the broadening functions as used for the 1H phase, provide a calculated spectrum that agrees with the experimentally obtained valence band spectrum of the metastable phase (see Fig. 4d). Such remarkable agreement between the experiment and calculation establishes that the chemically exfoliated, few layer MoS$_2$ forms in the insulating 1T′ 2a×a superstructure, thereby suggesting a reevaluation and reinterpretation of extensive previous reports attributing many beneficial properties of chemically exfoliated, few layered MoS$_2$ to the presence of metallic 1T phase.

In the above, we arrived at the geometric structure of the metastable phase via a comparison of the experimental valence band spectrum and state-of-the-art electronic structure calculations corresponding to different geometric structures. Using Raman spectra, we now provide an even more direct evidence of the specific 1T′ phase being the dominant metastable phase in chemically exfoliated MoS$_2$. We clearly see (Fig. 2b) that the metastable phase is characterized by the emergence of three new peaks (156, 227 and 330 cm$^{-1}$), not present in the 1H phase. Calculated vibrational frequencies[23] are 151, 223 and 333 cm$^{-1}$ for the 1T′ phase and 178, 270 and 376 cm$^{-1}$ for the 1T′′′ phase, whereas some phonon branches for the 1T structure was shown to have imaginary frequencies indicating this structure to be inherently unstable[25]. Thus, the vibrational frequencies observed with the formation of the metastable phase (Fig. 2b) are consistent only with the 1T′ 2a×a superstructure and in obvious disagreement with the expected frequencies of 1T′′′ structure. It is interesting to note that vibrational frequencies so far reported [23, 32, 42-47] for the chemically exfoliated $MoS_2$ samples are in close agreement with our results, thereby establishing that this route of exfoliation invariably results in the dominant presence of the 1T′ 2a×a superstructure of $MoS_2$ independent of the exact details of the preparation process. We note that there have been reports of direct structural investigations of chemically exfoliated $MoS_2$ suggesting the formation of other forms, including the metallic 1T phase, as already discussed in the introductory part of this report. All these investigations are based on high resolution TEM that looks at only a small nanoscopic fragment of the entire sample and therefore, such reports need not be in contradiction with our findings that establish that the 1T′ insulating 2a×a superstructure is the overwhelmingly dominant metastable phase formed when 1H $MoS_2$ is chemically exfoliated giving rise to form few layer $MoS_2$ samples.

In summary, we have unraveled the mystery of atomic and electronic structures of the metastable, trigonal form of 2D $MoS_2$, that occur in nano-scale regions of commonly grown samples and is known to have favorable properties for many applications. We used a combination of spatially resolved Photoemission spectroscopy of core electrons of Mo, to identify these regions precisely, and of valence electrons, in conjunction with self-energy corrected first-principles calculations of electronic structure to establish its electronic and geometric structures, corroborated further with

Raman spectroscopy. Thus, the observed metastable phase of 2D $MoS_2$ has a $2a \times a$ superstructure with Mo dimerization, and exhibits a band-gap of ~90 meV.

**Methods:**

**Chemically exfoliated $MoS_2$ synthesis:** Extensive occurrence of the minority phase of $MoS_2$ was stabilized on the indium doped tin oxide (ITO) substrates using the well-known and standardized organolithium chemistry. Mechanically exfoliated flakes of 1H $MoS_2$ deposited on the ITO substrate were treated with 1.6 M n-butyl lithium in hexane solution for two hours in argon atmosphere at the room temperature (300 K). After the completion of the reaction, the removal of unreacted n-butyl lithium from the sample was ensured by several washes with hexane. Finally, the ITO substrate containing the sample was washed with distilled water to remove any excess intercalated Lithium ions.

**Raman Measurement:** Micro-raman measurements were carried out in ambient condition and at room temperature. First with the help of an optical microscope the patches of $MoS_2$ was located. After that a green Laser beam was focused at the middle of one selected patches and the raman spectra were collected.

**SPEM measurements:** We have performed Scanning Photo Electron Microscopy (SPEM) at to understand the electronic structure of the metastable phase. This technique involves XPS studies with a highly focused photon beam achieved using Fresnel zone plates (ZP) on a high brilliance beamline at a third generation synchrotron facility with the sample mounted on a piezo driven scanning stage. Our experiment was performed at ESCA Microscopy beamline of Elettra Sincrotrone Trieste, with $h\nu = 400$ eV, energy resolution of 180 meV and spatial resolution of ~120 nm. All the SPEM measurements were carried out at room temperature at with a chamber pressure ~5×10$^{-10}$ mbar.

**Computational methods:** We show the partial densities of states for Mo 4*d* and S 3*p* based obtained from first-principles calculations based on density functional theory as implemented in Vienna *ab initio* package (VASP)[37]. Core and valence electrons are treated using the projected-augmented wave (PAW) method[38, 39]. We used the GW version of PAW potentials available with VASP, which are expected to provide improved scattering properties at high energies[40]. The exchange-correlation of electrons is treated using a generalized gradient approximation with functional parameterized by Perdew, Burke, and Ernzerhof (PBE). We use an energy cutoff of 380 eV to truncate the plane wave basis used in representing Kohn-Sham wave functions. We relaxed the structures to minimize energy until the Hellman- Feynman force on each atom is less than 0.01 eV/Å. We have used a periodic supercell to simulate a 2D MoS$_2$ sheet, with a vacuum of 10 Å to separate the periodic images. Integrations over the Brillouin zone (BZ) were sampled on 24×24×1 and 12×24×1 uniform meshes of k-points (Γ-centered) for 1H and 1T´ (2a×a) respectively. Converged PBE eigenvalues and wave functions were used subsequently to calculate quasiparticle energies in the non-self-consistent GW (G0W0) approximation.

**Figure captions:**

**Figure: 1.** Top view of MoS$_2$: 1H, 1T and various distorted 1T structures (1T´, 1T´´ and 1T´´´). 1T´ can be described in terms of two different unit cells corresponding to 2a×a and √3a×a supercells. The unit cells are enclosed by solid lines.

**Figure: 2.** Room-temperature Raman spectra from different spots on pure 1H (sample A) and mixed phase (sample B). (a) sample A, and (b) sample B.

**Figure: 3.** Spectral features and intensity maps of Mo 3d region from samples investigated. (a) Mo 3$d$ spectra of samples A (filled red circles) and B (black open circles). (b) Map of Mo 3$d_{5/2}$ signal intensity over the entire energy range of 228-230 eV in sample B. (c) Map of the intensity ratio, I/II, over respective energy windows shown in panel a for the total intensity map in (b). (d) Mo 3$d$ spectrum (open circles) from sample B and the fitting of this spectrum (solid line) with two contributions (dashed and dotted lines). S 2$s$ contribution is also shown.

**Figure: 4.** Experimentally and theoretically obtained valence band spectra from pure 1H (sample A), mixed phase (sample B) and contributions from the metastable phase alone. (a) Experimental (open circle) and calculated (solid line) valence band spectra of pure 1H sample along with partial densities of states of Mo 4$d$ (dash) and S 3$p$ (dot) states. (b) Valence band spectra of mixed phase (solid line) and pure 1H (circles) MoS$_2$ near the Fermi energy. Inset shows the valence band spectrum of the mixed phase sample over a wide energy range. (c) Contributions from pure 1T phase (triangles) and the substrate (solid line) were subtracted from the total valence band spectrum of the mixed phase sample to obtain the spectral features (open circles) of the metastable phase. (d) Experimentally obtained spectral features of the metastable phase (open circles) compared with that obtained from the calculation of the electronic structure of the metastable 1T´ phase of MoS$_2$ (solid line) along with partial densities of states of Mo 4$d$ (dash) and S 3$p$ (dot) states.

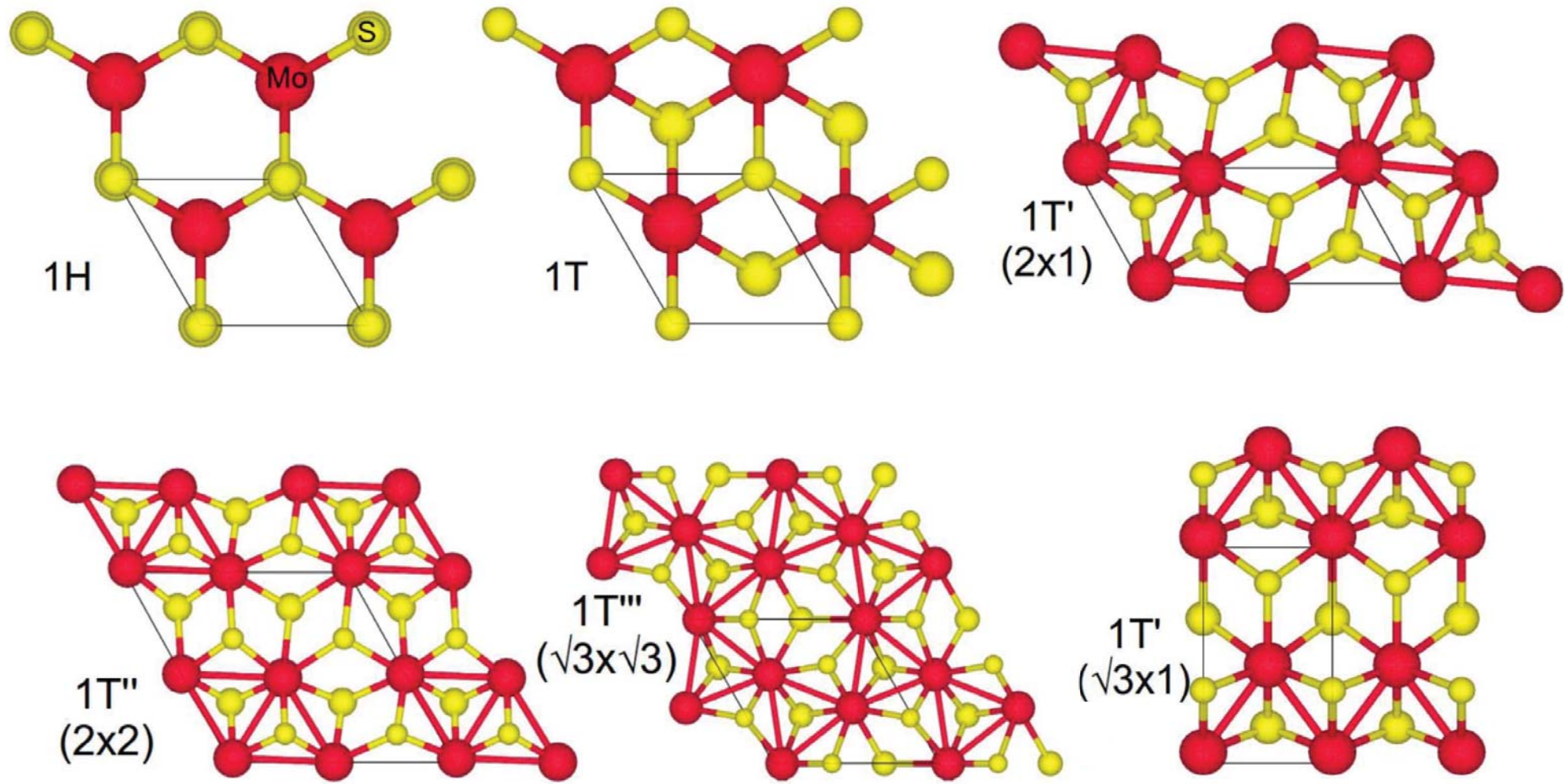

Figure: 1

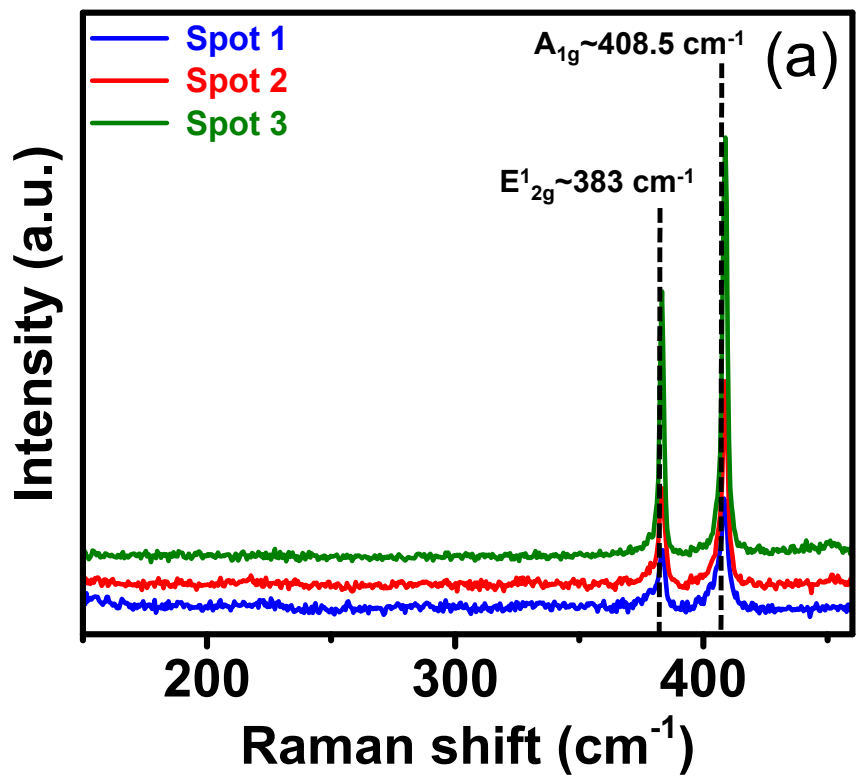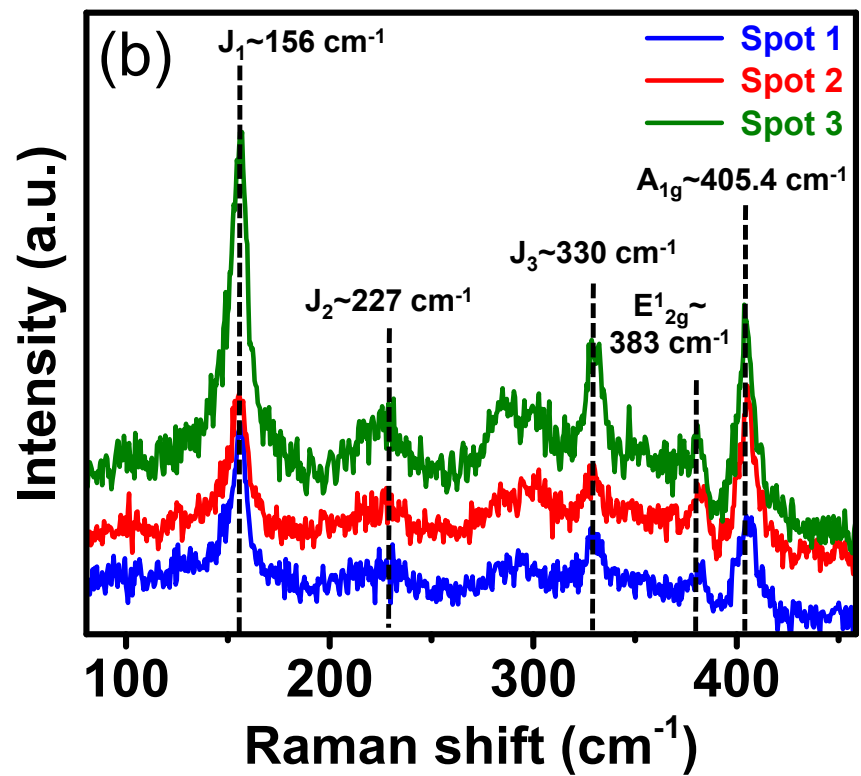

Figure: 2

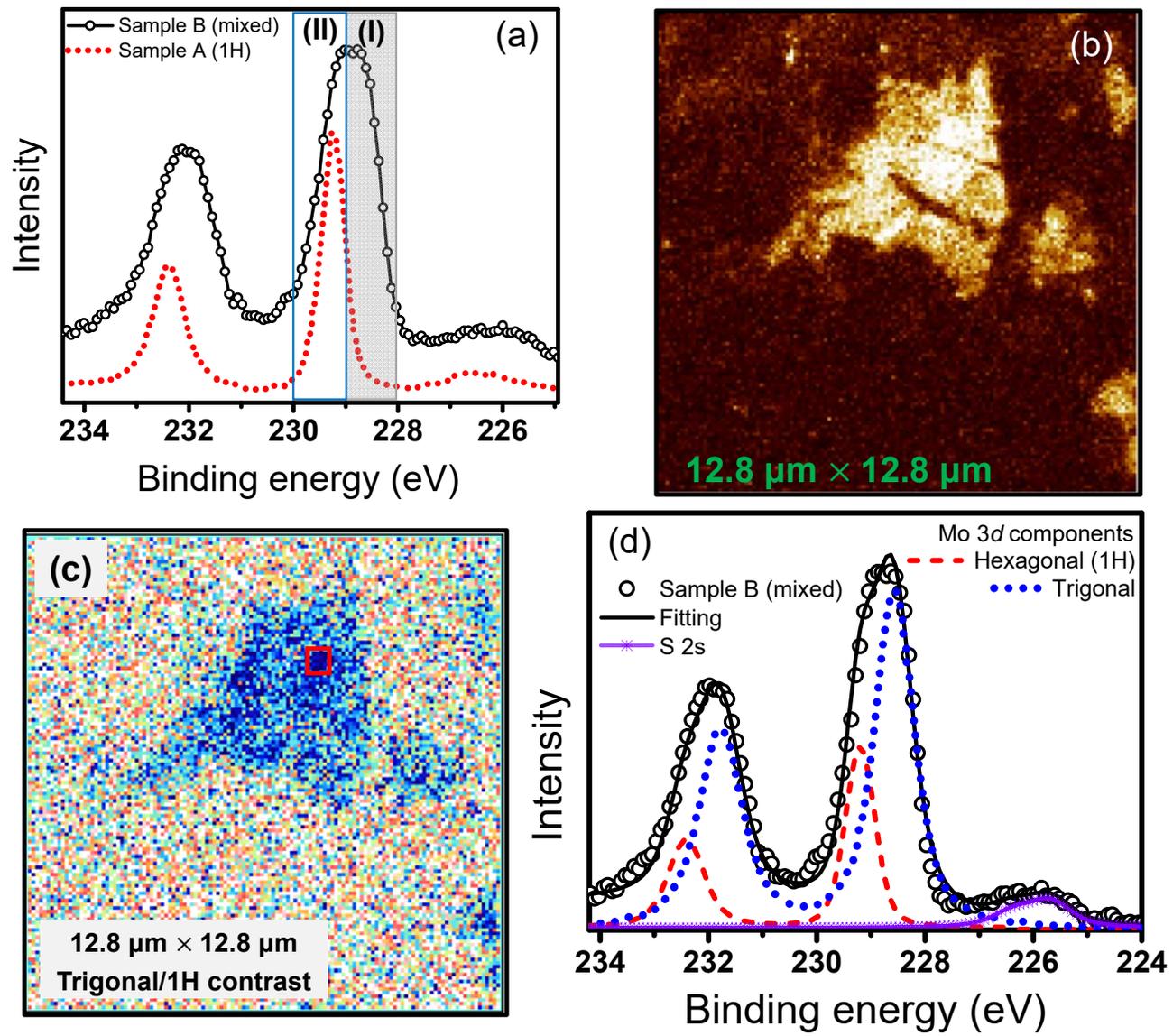

Figure: 3

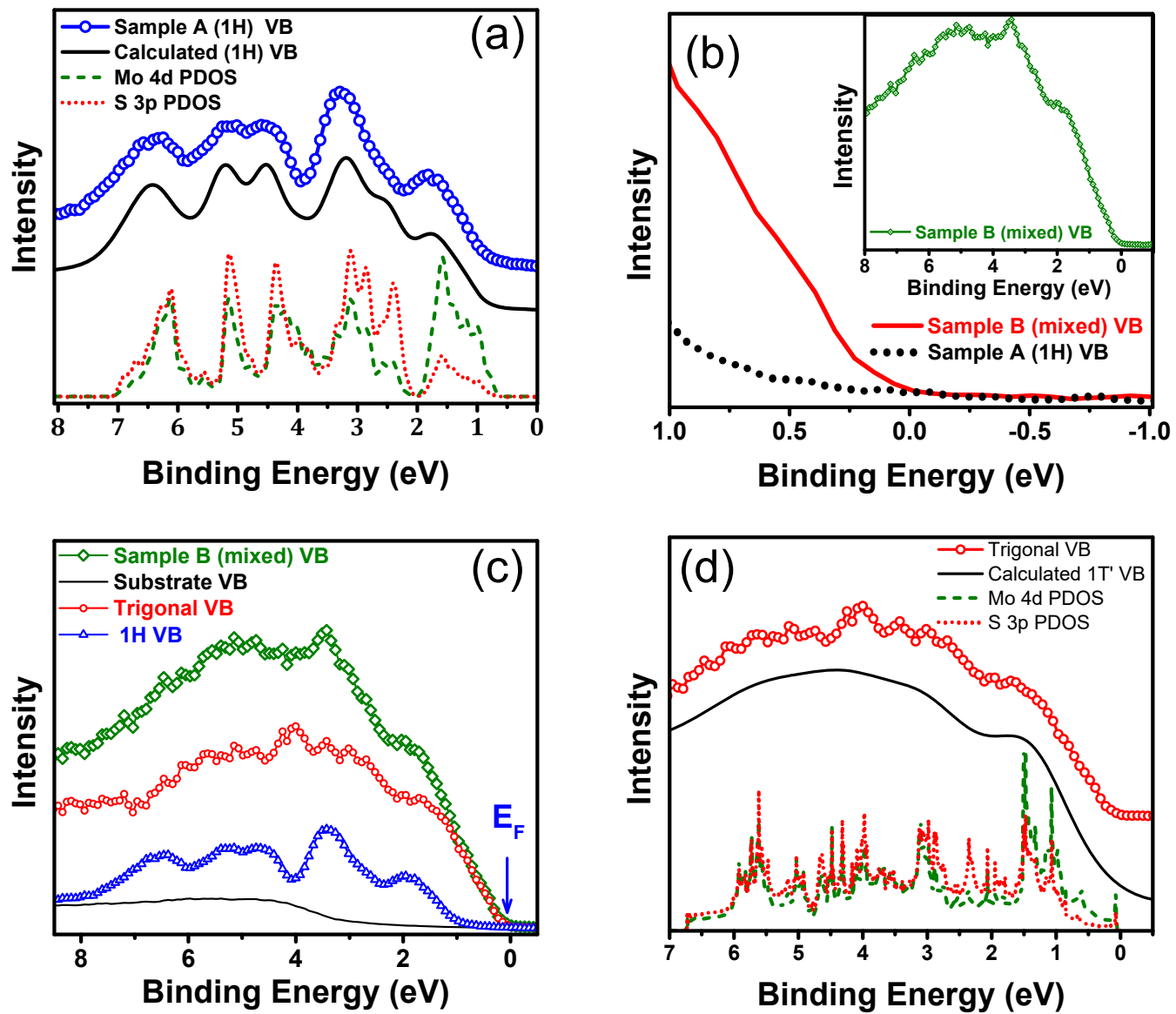

Figure: 4